\documentclass[aip, rsi, reprint, floatfix]{revtex4-1}

\usepackage{graphicx}
\usepackage{dcolumn}
\usepackage{bm}
\usepackage{amsmath}

\usepackage[utf8]{inputenc}
\usepackage[T1]{fontenc}
\usepackage{mathptmx}
\usepackage{etoolbox}
\usepackage{gensymb}
\usepackage{xcolor}

\def\persec{\mathrm{s^{-1}}}

\def\2D{\mathrm{2D}}
\def\3D{\mathrm{3D}}

\def\Wpermm{\mathrm{W \cdot m^{-2}}}
\def\kBT{k_B T}

\def\rhounits{\mathrm{kg \cdot m^{-3}}}
\def\cmrunits{\mathrm{C \cdot kg^{-1}}}
\def\speedunits{\mathrm{m\cdot s^{-1}}}

\def\Vac{V_{\mathrm{ac}}}
\def\Vdc{V_{\mathrm{dc}}}
\def\VAC{V_{\mathrm{ac}}}

\def\Wac{\Omega}

\def\N2{$\mathrm{N}_2$}

\def\CMR{Q/M}

\def\AmCa{$(\mathrm{NH}_4)_2\mathrm{CO}_3~$}

\def\h2o{$\mathrm{H}_2\mathrm{O}$}
\def\co2{$\mathrm{CO}_2$}
\def\c2h2{$\mathrm{C}_2\mathrm{H_2}$}
\def\o2{$\mathrm{O}_2$}

\begin{document}


\title{\begin{center}Collection, characterization, and precision measurement of levitated charged nanoparticles
    \end{center}} 



\author{B. E. Kane}
    \email{bekane@umd.edu}
    \affiliation{Laboratory for Physical Sciences, 8050 Greenmead Dr., College Park, MD, 20740, USA}
    \affiliation{Joint Quantum Institute, University of Maryland, College Park, MD, 20742, USA} 
 \author{Joyce Coppock}
  \affiliation{Laboratory for Physical Sciences, 8050 Greenmead Dr., College Park, MD, 20740, USA}
    \affiliation{University of Maryland, College Park, MD, 20742, USA}
   
\author{Sunghyun Kim}
   
    \affiliation{Laboratory for Physical Sciences, 8050 Greenmead Dr., College Park, MD, 20740, USA}
     \affiliation{University of Maryland, College Park, MD, 20742, USA}
 \author{Sarah Westgate}
  \altaffiliation[Current address: ]{University of Bordeaux, CNRS, LOMA, UMR 5798, F-33400 Talence, France}
  \affiliation{University of Maryland, College Park, MD, 20742, USA}


\date{\today} 

\begin{abstract}
We describe apparatus and experimental procedures for high stability precision measurements of levitated nanoscale particles confined in an ion trap in high vacuum.  We discuss methods for particle generation and collection using electrospray emission, for rapid characterization by direct imaging of thermal motion, and for transfer of the particle from the trap where it is collected to a separate analysis trap in order to achieve better vacuum and lower noise. In the analysis trap at high vacuum (pressure $p\simeq10^{-8}$ Torr), we employ thermostatic control of the trapped particle oscillation amplitudes, allowing long-term, precision measurements of oscillation frequencies, from which the charge to mass ratio ($\CMR$) can be deduced. Under these conditions, we achieve $\CMR$ measurement precision approaching $10^{-5}$. This sensitivity will enable, for example, investigations of the surface chemistry  of $\mu$m-scale levitated materials in ultra high vacuum environments.
\end{abstract}


\maketitle 


\section{Introduction}
\label{sec:intro}
Levitation of nanoscale materials in optical tweezers or quadrupole ion traps has many applications, ranging from quantum sensing to materials science \cite{Gonzalez2021}. In probes of the material science and surface chemistry of nanoscale objects, important information can be gleaned from careful measurements of the mass of the object. Data on surface adsorption and desorption \cite{Hoffman2020} have been collected that have been used to probe surface plasmon resonances of Au nanoparticles\cite{Hoffman2023} and temperature dependent changes in levitated silica \cite{Ricci2022}. Sublimation and  oxidation have been measured at extremely high temperatures of carbon\cite{Long2020,Rodriguez2021} and refractory metals\cite{Friese2025}. Also, growth of C materials has been observed in a \c2h2 ambient\cite{Lau2023}. Careful mass measurements have also been used to probe melting\cite{Coppock2021} and supercooling\cite{Coppock2022} of levitated Au nanoparticles.

Mass ($M$) for particles confined in ion (Paul) traps is determined from the charge to mass ratio ($\CMR$) which in turn is inferred from particle oscillation frequencies ($\nu_x,\nu_y,\nu_z$) in the trap.  Recently, quality factors in excess of $10^{10}$ have been reported for particles oscillating in ion traps in an ultra-high vacuum (pressure $p=7\times10^{-11}$ mbar) environment\cite{Dania2024}. In these experiments, Allan deviation for the oscillation frequency reached a minimum of $2\times10^{-6}$, suggesting that a similar precision is possible for measurements of $\CMR$ and $M$ in high vacuum. 

We describe below the experimental instrumentation and techniques that we have developed that are optimized for high-stability precision measurements of particles confined in ion traps in high vacuum ($10^{-8}\leq p\leq10^{-5}$ Torr). After a discussion of the electrospray method we use to generate and collect particles (Sections \ref{sec:generation}-\ref{sec:trap}), we present a technique (Section \ref{sec:characterization}) for characterization ($M$ and radius $R$) of the particle by rapid imaging\cite{Minowa2022} of its thermal (Brownian) motion. These measurements also provide insight into the excess noise that originates from charge on the surfaces of trap electrodes\cite{Pontin2020}: particle injection techniques like electrospray--which deliver charged particles not only to the trap, but also deposit them on the nearby electrodes--create excess noise that slowly decays (over days) until thermal equilibrium is reached.

\begin{figure}
\includegraphics[scale=1.0,draft=false]{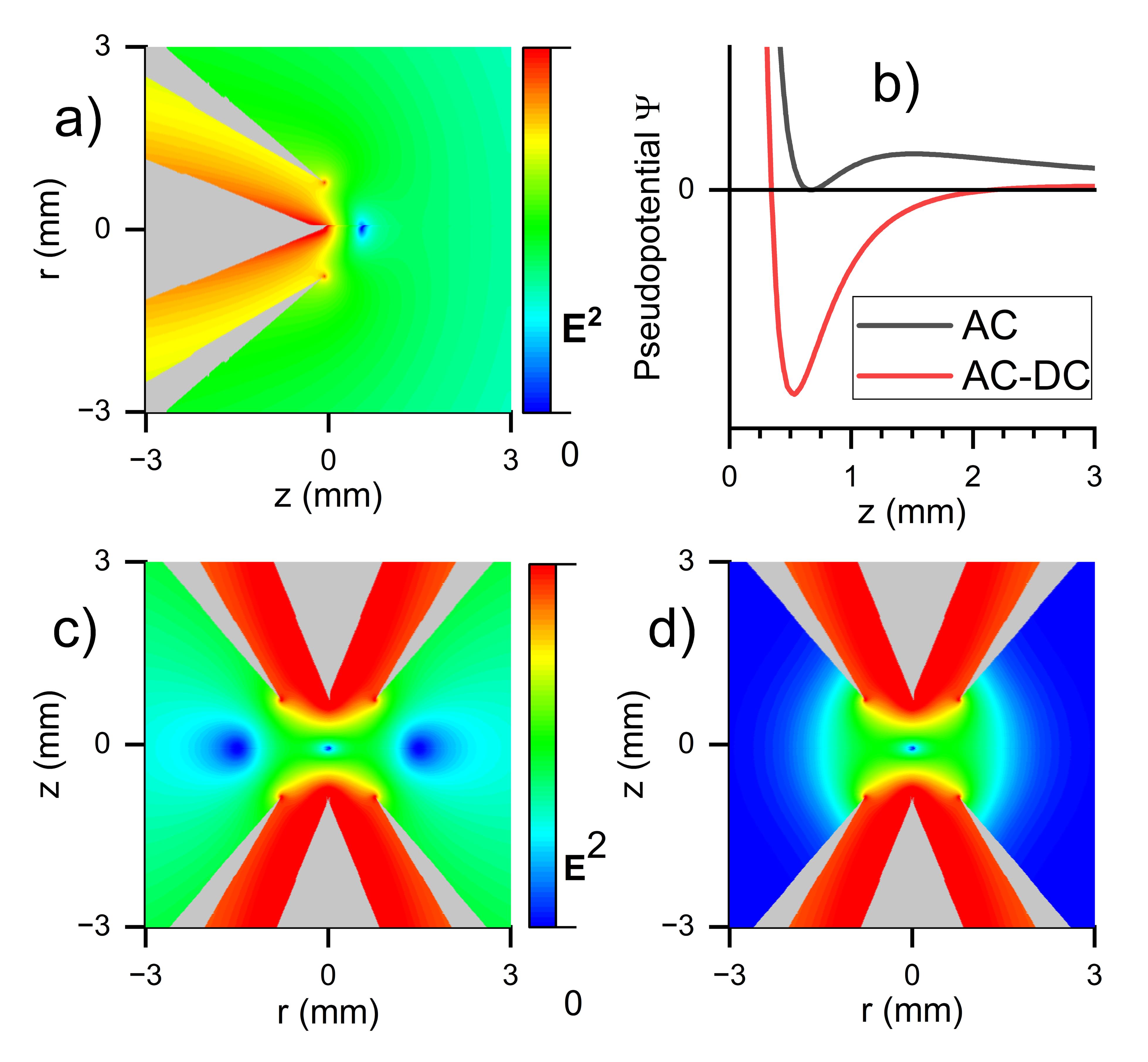} 
\caption{
(a) COMSOL\cite{COMSOL} model of the pseudopotential (see Eq. \ref{Psi}) of the collection trap, in which the inner electrode is grounded and a bias of $\Vac$ is applied to the outer electrode. (b) When a constant voltage $V_{dc}$ (negative for positively charged particles) is applied to the inner electrode, the location of the trap potential minimum moves toward the trap face and the volume of the basin of attraction greatly expands, allowing for accelerated capture of charged particles in the vicinity of the trap. (c,d) Symmetric arrangement of two pairs of electrodes, each identical to the single trap shown in (a), with $\Vac$ applied to the outer (c) and inner (d) electrodes, each with alternate electrodes grounded. A trap potential minimum lies at the center of both configurations, but in the case where the outer electrodes are biased (c), an additional toroidal minimum or halo arises, due to the boundary conditions imposed by the conducting enclosure at ground (not shown).
}
\label{COMSOL}
\end{figure}

To completely remove the particle from the environment where it is collected and move it into an environment optimized for precision measurements, we describe our process for transferring particles between two ion traps (Section \ref{sec:transfer}).  We have previously reported on this technique\cite{Coppock2017} using two traps, each consisting of a single pair  of coaxial conically shaped electrodes (Fig. \ref{COMSOL}a) oriented on perpendicular axes. We show that it  is also possible to transfer into a symmetric arrangement of two coaxial conical electrode pairs facing each other (Fig. \ref{COMSOL}c,d), a trap geometry similar to those adopted by several groups \cite{NisbetJones2016,Hoffman2020,Lindvall2022}.  In this configuration there is almost 360$\degree$ access for optics and beams in the symmetry plane, and the symmetric design means that the achievable electric field strengths at the trap center (for example, for coupling to particle electric dipoles \cite{Nagornykh2017}) are approximately double those possible with our previous design.

The remainder of the paper focuses on techniques for making stable, long term, precision measurements of $\CMR$ in high vacuum conditions, where relaxation rates due to interactions with residual gas are negligible. In Section \ref{sec:Control}, we discuss a method for using three phase locked loops (PLLs) to track trapped particle motion along all three axes of confinement.  Using these PLL outputs, we implement thermostatic control of particle motion in order to maintain amplitudes that are optimal for measuring motional frequencies and determining $\CMR$. Using these techniques, we  have been able to lock and stabilize particle motion for over a month.

Finally, in Section \ref{sec:precision}, we present a discussion of  our methods for collecting and  analyzing $\CMR$ data on an Au nanoparticle at $p\sim10^{-8}$ Torr. We report a peak precision of  $\delta(Q/M)/\langle Q/M \rangle\cong10^{-5}$ for data averaged over ten minute periods. Au, like many other materials of interest, is highly absorptive, so our measurements are performed using an illumination of only 1000 $\Wpermm$ at 532 nm.  This power density is two orders of magnitude less than that used on silica nanospheres in Ref.\citenum{Dania2024}, Nevertheless, the precision reported in those results is only a factor of five better than what we have obtained for particles of similar dimensions.

\section{Particle Generation}
\label{sec:generation}

The apparatus used for particle generation and collection is shown in Fig. \ref{CollectionApparatus}a and is similar to what we have described previously \cite{Coppock2017,Coppock2021}. Au nanosphere suspensions are prepared  in the same way as described in Ref.\citenum{Coppock2021}. Commercial polystyrene nanosphere suspensions \cite{PSSource} are diluted (30:1 diluent:stock solution) with the same fluid we have used for Au (\AmCa diluted to a concentration of 2 mM with 1:3 isopropyl alcohol:deionized water) without any step to remove additives in the stock solution.

The electrospray capillary emitters we employ have a 100 $\mu$m inner diameter, were sourced commercially \cite{Emitters}, and are subsequently coated with a 25 nm layer of Pt using a vacuum coater. Emission occurs in a $\sim$10\%  \co2 90\% \N2 ambient at atmospheric pressure \cite{TSIApplicationNote}, typically at a suspension flow rate of 1 $\mu$l/minute and at an emitter bias of $V_e$= +2400 V (a condition that ensures that the particles we collect will have positive charge).  For most of our work, the emitter is located about 5 mm above the circular aperture \cite{Apertures} that separates the emission chamber from the evacuated trap chamber. The distance between the emitter tip and the aperture plate can be substantially reduced  if the aperture plate is heated to prevent the formation of droplets that can occlude the aperture.

\section{Particle acceleration and deceleration}
\label{sec:acceleration}

The inner diameter of the pinhole aperture that separates the electrospray emitter from the trap vacuum chamber should be large enough to avoid clogging and admit as many particles as possible.  However, the pressure in the trap region must be lower than $\sim$1 Torr to prevent corona discharge arising from the high voltages ($\Vac\simeq300$ V) on the trap electrodes.  For the pumping speeds available in our apparatus, 100 $\mu$m inner diameter pinhole apertures satisfy these conditions.  Standard calculations \cite{LenoxLaser} predict a peak \N2 gas flow velocity at an aperture of this size ($r_0$ = 50 $\mu$m) of $u_0\simeq$ 187 $\speedunits$.  Since particles created by electrospray will be accelerated to high velocities by this flow, it is important to approximate the necessary distance between the aperture and the trap to ensure that particles are slowed and come to a stop prior to capture.

\begin{figure}
\includegraphics[scale=1.00,draft=false]{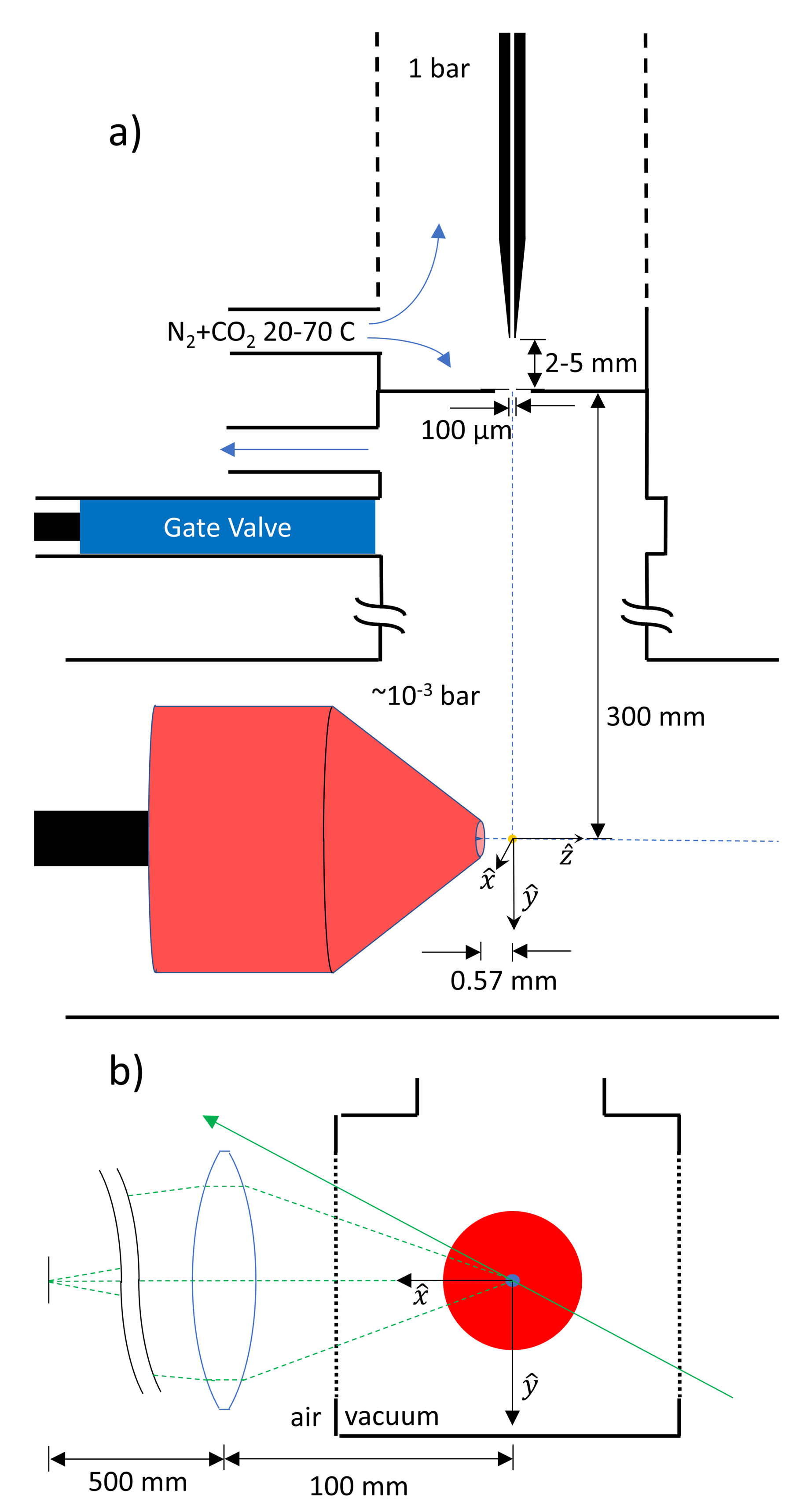} 
\caption{
(a) Particles are generated for collection in an ion trap by a Pt coated electrospray emitter  in a 90$\%$ \N2-10$\%$ \co2 ambient at atmospheric pressure, located above a pinhole aperture through which particles must pass in order to enter the partially evacuated trap chamber. Typical emission parameters are a flow rate of 1 $\mu$l/minute at an emitter potential of +2400 V. (b) View along the trap axis: the particle is illuminated by a 532 nm laser and imaged by a 50 mm diameter lens located outside the vacuum chamber.
}
\label{CollectionApparatus}
\end{figure}

\begin{figure}
\includegraphics[scale=1.00,draft=false]{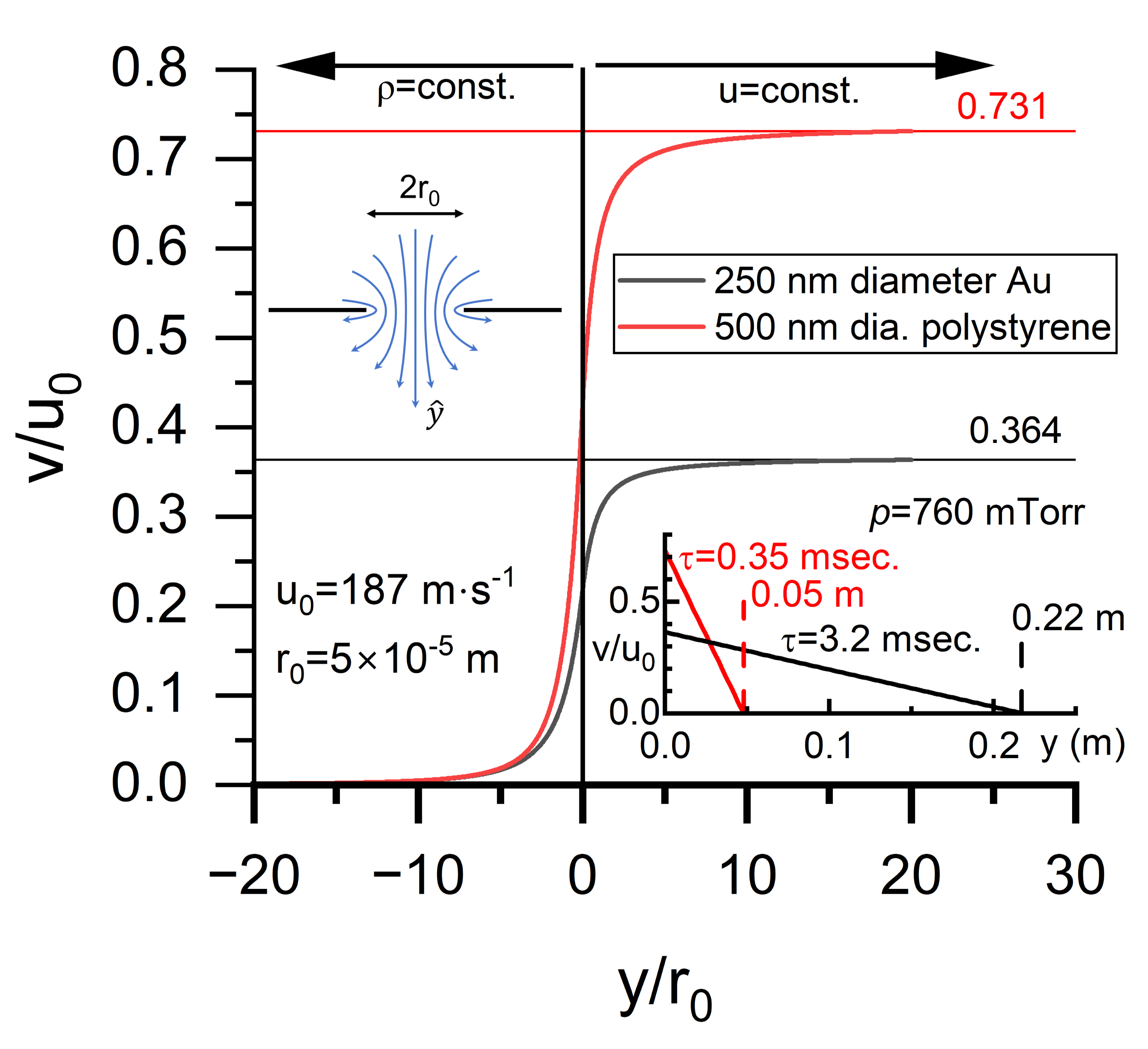} 
\caption{
Simple model of the acceleration and deceleration of a particle passing from a region at atmospheric pressure, through a circular aperture (at $y$ = 0), and in to a partially evacuated region where the trap is located. Acceleration occurs over distances comparable to the aperture radius, $r_0$, and the particle reaches velocities comparable to that of the gas at the flow constriction. Deceleration occurs much more slowly (inset), and dense particles like Au have a stopping distance $\sim$200 mm.
}
\label{AccelDecel}
\end{figure}

Perhaps the simplest model of gas flow through the aperture (Fig. \ref{AccelDecel}) is to posit a constant gas density, $\rho=\rho(\mathrm{STP})$, upstream of the aperture, a constant velocity, $u=u_0$, downstream of the aperture, and a  uniform mass flux density $=\rho(\mathrm{STP})u_0$ within the aperture:

\begin{equation} \label{Eq. a}
    \overline{\rho} =\frac{\rho (y)}{\rho(\mathrm{STP})}=
\begin{cases}
1, & y<0 \\
1-\sqrt{\frac{(y/r_0)^2}{1+(y/r_0)^2}}, & y\ge 0
\end{cases}
\end{equation}
and:
\begin{equation} \label{Eq. b}
    \frac{u(y)}{u_0}=
\begin{cases}
1-\sqrt{\frac{(y/r_0)^2}{1+(y/r_0)^2}}, & y<0 \\
1 & y\ge 0.
\end{cases}
\end{equation}
Here, $\rho(\mathrm{STP})$ is the density of the gas at atmospheric pressure and $T=293$ K. The $y$-axis is directed along the direction of particle motion (Fig. \Ref{CollectionApparatus}a) with $y$ = 0 in the center of the pinhole aperture. These equations are appropriate for idealized laminar flow through the aperture.  Solid particles in this flow will experience a force\cite{Li2003} :

\begin{equation} \label{Eq. c}
\mathbf{F} = -\frac{6 \pi \mu R \mathbf{V}}
{1 + \mathrm{Kn}\left[A + B \exp\left(-\frac{E}{\mathrm{Kn}}\right)\right]},
\end{equation}
where $\mathbf{V}$ is the velocity of the particle relative to the surrounding gas.
Here $\mu$ is the viscosity, $R$ is the particle radius, and $\mathrm{Kn}$ is the Knudsen number = $\lambda/R$, with $\lambda$ the molecular mean free path. $A$=1.155, $B$=0.471 and $E$=0.596 are constants that have been determined from experiments in air \cite{Li2003,Allen1982}. For \N2, $\mu$ = 1.78$\times10^{-5}$ Pa$\cdot$s and $\lambda$ = 6.87$\times 10^{-8}$ m/$~\overline{\rho}$.

We performed numerical integration of the above equations to determine the asymptotic velocities for two types of test particles: spherical Au with diameter $D$ = 250 nm and $\rho_{\mathrm{Au}}$ = 19,300 $\rhounits$ ; also, polystyrene with $D$ = 500 nm and $\rho_{\mathrm{PS}}$ = 1050 $\rhounits$. As shown in Fig. \ref{AccelDecel},  the particles accelerate to a velocity somewhat less  than $u_0$, with the Au particle significantly slower, due to its greater inertia. While particle acceleration occurs near the aperture, deceleration occurs much more slowly, due to the much lower gas pressure in the collection chamber when $y/r_0\gg1$ (Fig. \ref{AccelDecel}, inset). In this regime, we assume that the residual gas is at rest  relative  to the particle. Again, velocity change is less rapid for the denser Au particle, which takes over 20 cm to  come to rest.

This model has obvious deficiencies: fluid dynamics and the effects of charge on the particles are neglected.  Also, the complexities of mass loss of the particles as liquid evaporates or droplets fission are not considered. We nonetheless expect that the model provides crude estimates of typical particle velocities and stopping distances that are not too far off.

\section{Particle Trapping}
\label{sec:trap}

Particle trapping  occurs in a trap located about 300 mm below the pinhole aperture.  The trap consists of two stainless steel coaxial conical electrodes (Fig. \ref{COMSOL}a, Fig. \ref{CollectionApparatus}a) that are similar in design to what we have reported on previously\cite{Kane2010,Coppock2017}.  For trapping, the outer (ac) electrode is typically biased at $\Vac$ = 300 V at a frequency $\Wac/2\pi\sim$3-50 kHz, while the inner (dc) electrode is biased at $\Vdc$ = -10-0 V. The pseudopotential\cite{Kane2010} at low pressures is given by:

\begin{equation} \label{Psi}
\Psi = \frac{1}{4}\frac{Q}{M}\frac{1}{\Wac} E^2_\mathrm{ac}+\phi_\mathrm{dc}.
\end{equation}
$E_\mathrm{ac}$ is the electric field derived from electrostatics calculations for $\Vac\neq$0 and the inner electrode grounded. $\phi_\mathrm{dc}$ is determined for the condition when $\Vdc\neq$0 and the outer electrode is grounded.  In both cases, the surrounding vacuum chamber is an additional grounded surface.
When $\Vdc$ = 0, $\Psi$ has a minimum located about 0.66 mm from the apex  of the inner electrode. Applying $\Vac$ and a simultaneous $\Vdc$ with polarity opposite to the particle charge creates a greatly expanded basin of attraction for the trap that increases the rate of collection of charged particles passing by (Fig. \ref{COMSOL}b).

Particle capture in the ion trap is observed by imaging the particle using an illumination laser ($\lambda$ = 532 nm), a lens outside the vacuum chamber, and a camera (Fig. \ref{CollectionApparatus}b). Laser power is typically 30-100 $\mu$W for a beam with diameter $\sim$1 mm. Due to the term $\CMR$ in Eq. \ref{Psi}, a negative $\Vdc$  applied during collection will have the greatest effect on particles with the smallest $\CMR$, so a crude estimate of $\CMR$ is possible immediately after particle capture by viewing the particle location in the image. 

As will be discussed further in the following section, charged droplets reaching the trap during electrospraying can lead to excess surface charge on the electrodes. Consequently, we open the gate valve that separates the emission source from the trap (Fig. \ref{CollectionApparatus}a) only after a stable plume of the particle suspension has been established. After the gate valve is opened, typically 0.1-10 objects/second are observed being collected in the trap, using the suspensions and emission  parameters discussed above.  Almost all of these are liquid droplets that rapidly lose mass and are quickly (seconds) ejected from the trap.   Solid particles are collected much less frequently ($\sim$ every several minutes), with the ratio of droplet  to solid particle collection rates comparable to the proportion of solid particles in the suspension. Unlike the case for droplets, solid particle $\CMR$ does not change significantly after the moment of collection, but its value can vary widely from $\sim$0.1-10 $\cmrunits$.

\section{Rapid characterization of collected particles}
\label{sec:characterization}

From a slow camera (frame rate $\ll \nu_x, \nu_y, \nu_z$), only two parameters of the particle are easily determined: the particle brightness (which provides a crude measure of its size) and $\CMR$, inferred from the particle's sensitivity to changes in $\Vdc$. To obtain more extensive data on the particle soon after it is collected, we have developed a technique that performs rapid imaging\cite{Minowa2022} on the particle to observe thermal motion in the trap. At sufficiently low pressure and when trap parameters are far from instability, particle motion is nearly harmonic in the three dimensional trap potential.  If the particle is in thermal equilibrium with the surrounding gas at temperature $T$, then the mass can be determined from the following:

\begin{equation} \label{Eq. e}
M=\frac{\kBT}{\omega_z^2 \langle z^2 \rangle},
\end{equation}
where $\langle z^2 \rangle$ is the mean square displacement along the $z$ axis of the particle measured from the trap center and $\omega_z=2 \pi \nu_z$ is its angular velocity of oscillation. (A similar equation applies to the motion in the $y$ axis direction.) Applying Eq. \ref{Eq. c} to the conditions at low pressure when Kn$\gg$1, we get an expression for the relaxation rate of motion in the trap:

\begin{equation} \label{Eq. f}
\frac{1}{\tau}=\frac{6\pi\mu R^2}{1.626 M \lambda}=2\pi \Delta \nu
\end{equation}
If $p$ is known, the fact that $p \lambda$ is constant and $\mu$ is independent of $p$ enables us to determine the radius of a trapped spherical particle from observations of its motion.

Measurements are performed with a camera \cite{Lucid} capable of taking data at 1000 frames per second for a 128 $\times$ 128 pixel region of interest (ROI). The camera pixel size is 6.9 $\mu$m and the magnification of the collection optics is 5 (Fig. \ref{CollectionApparatus}b). For each ROI frame, the $y$ and $z$ pixel locations in the image, weighted by the number of counts in the pixel, are averaged to determine the mean position of the particle: ($\langle y\rangle,\langle z \rangle$). Using this data, separate $\langle y\rangle$ and $\langle z \rangle$  fast Fourier transform (FFT) power spectra are calculated every 10 s, and the result is appropriately scaled for rectangular windowing. Thus, oscillatory motion along both the $y$ and $z$ axes at frequencies up to 500 Hz are measured with a resolution of 0.1 Hz. 

\begin{figure}
\includegraphics[scale=1.00,draft=false]{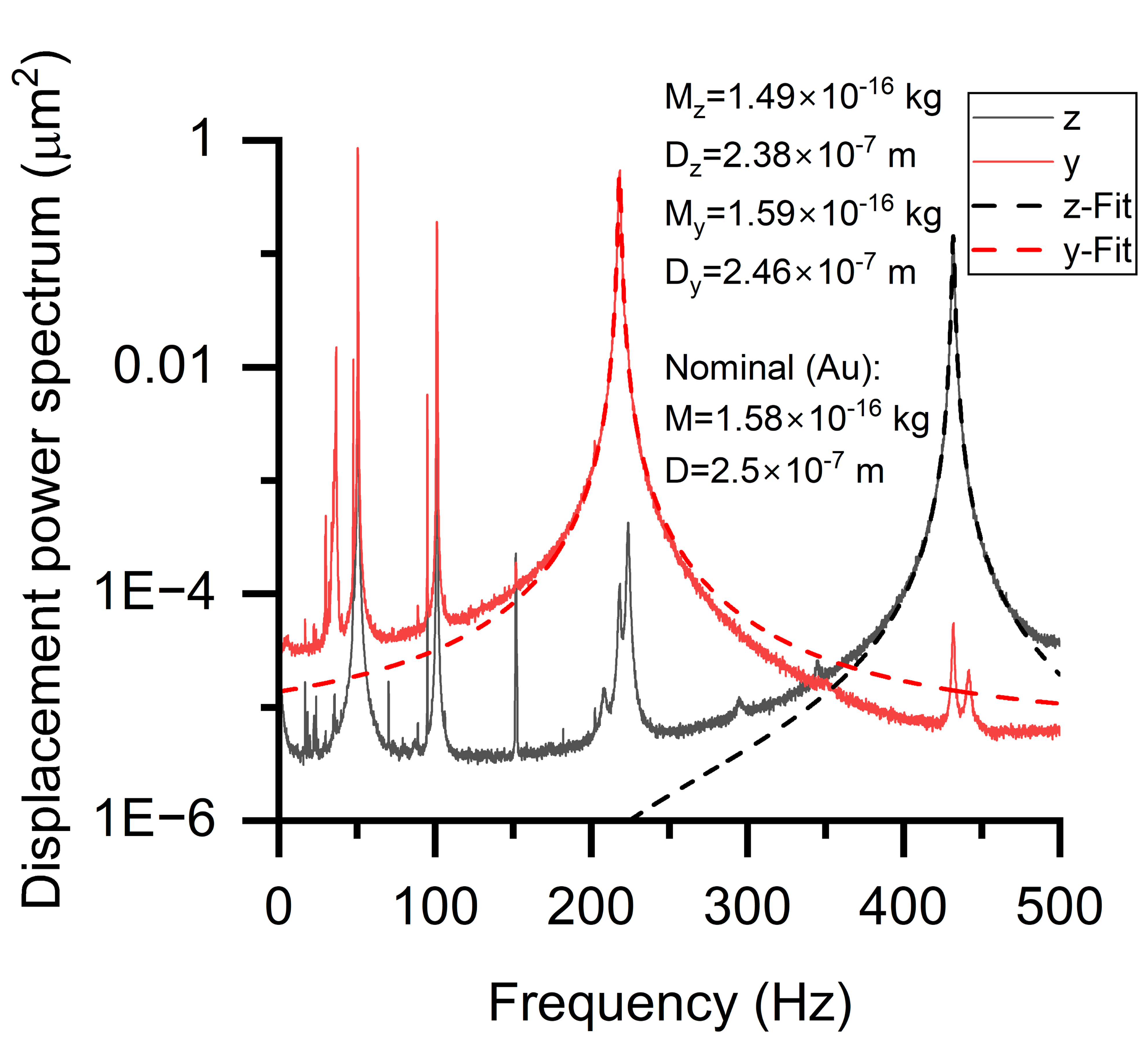} 
\caption{
Displacement power spectrum determined from camera images and the known optical magnification. Particle location along the trap axis ($z$) and perpendicular to it ($y$) are both determined at a rate of 1000 Hz, so the trap frequency $\Wac/2 \pi$ must be adjusted so that both $\nu_z$ and $\nu_y$ are at frequencies <500 Hz.  Sharp lines appearing at <200 Hz are from pumps and instrumentation.  Data was taken approximately four days after particle collection at $p=$25 mTorr and with laser power $=$75 $\mu$W.
}
\label{Thermal}
\end{figure}

While particles are collected at $p\cong$700 mTorr, optimal data for spectra for particles of sub-$\mu$m dimensions is obtained when $p=$10-30 mTorr. In Fig. \ref{Thermal}, data is shown for an Au nanosphere with nominal diameter $D=$250 nm. In the absence of dc field gradients, oscillation frequencies in the trap potential should  approximately obey the condition: $\nu_x+\nu_y=\nu_z$ (see Section \ref{sec:precision}).  Furthermore, because the potentials associated with the collection trap are symmetric around the $z$ axis, $\nu_x=\nu_y$, and thus $\nu_z=2\nu_y$.  Note that in the geometry we use to image particles in the collection trap, we cannot measure motion in the direction of the  $x$-axis.

From these considerations, we expect to observe a peak for motion along the $z$-axis at a frequency twice that of a peak in the $y$-axis data.  From Eqs. \ref{Eq. e} and \ref{Eq. f} we expect the $z$-peak to have a smaller magnitude than the $y$-peak, but both peaks should  have similar widths.  That is exactly what is seen in Fig. \ref{Thermal}, and fits to the spectra produce values of $M$ and $D$ within 10\% of the values expected from the manufacturer's specifications\cite{Cytodiagnostics} for the nanosphere dimensions.

Because $\CMR$ of the particles--and thus their oscillation frequencies--can vary substantially, it is necessary to adjust $\Wac$ so that both $y$ and $z$ peaks are found at <500 Hz. Aliased peaks can appear in the spectrum originating from oscillations at >500 Hz. Measuring the effect on peak positions from small changes in $\Wac$ can be used to distinguish between valid and aliased peaks. Finally, data must be taken in a regime when changes in laser power have no effect on the data, ensuring that the particles are in thermal equilibrium with background gas at $T\cong$293 K.

While we have established that high quality data like that shown in Fig. \ref{Thermal} is possible to obtain, we have found that to do so it is necessary to wait a substantial amount of time ($\sim$days) for peak amplitudes and positions to stabilize. When particles are collected during the electrospray process, charged particles inevitably also land on the trap electrodes.  Accumulation of these charges leads to electric fields, primarily along the $z$-axis, that slowly decay with time. While we are able to null $E_z$ at the trap center by applying a nonzero  $\Vdc$ to the inner electrode, both spatial and time derivatives of the electric fields will still impact the data.  We have found that immediately after a particle is collected, application of $V_{dc}=-1--2$ V is necessary to null stray fields emanating from the trap electrodes.  Over a few  days, the necessary nulling potential rises until $|V_{dc}|<0.5$ V.  During this time, peak positions shift, and their amplitudes, especially for motion along the $z$-axis, significantly exceed the values predicted from the assumption that the particle is in thermal equilibrium.

\begin{figure}
\includegraphics[scale=1,draft=false]{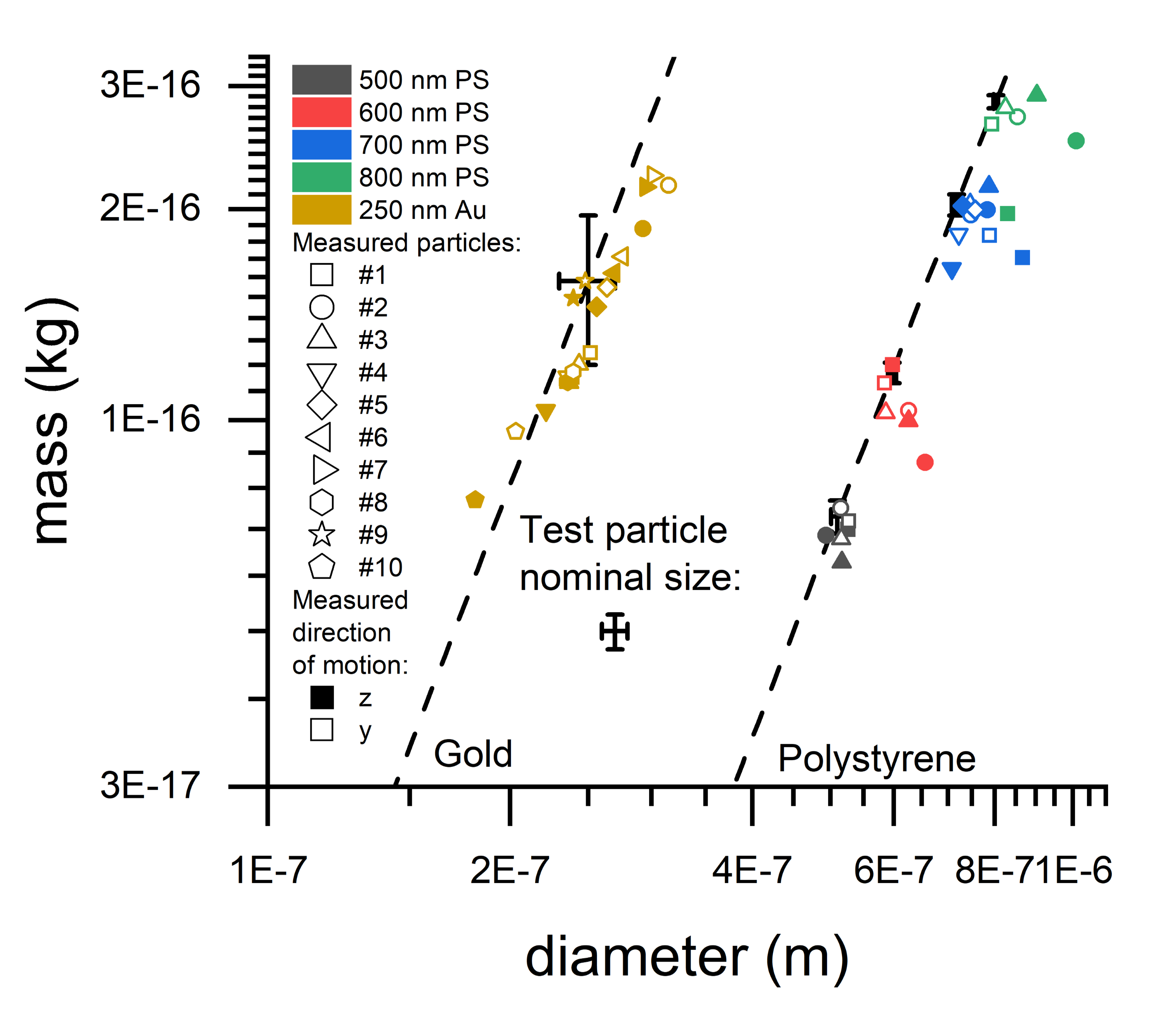} 
\caption{
Mass and diameter of Au and calibrated polystyrene nanospheres inferred from particle motion in an ion trap, under the assumption that they are in thermal equilibrium with the surrounding gas.  Variance crosses are determined from the manufacturer's specifications and the known density of the materials.  Densities are also used to compute the dashed lines. Data was taken with $p=$10-30 mTorr.
}
\label{Sarah}
\end{figure}

Extensive data for several test particles is presented in Fig. \ref{Sarah}.  In almost all cases, data was taken within 24 hours of collection.  Data for $z$-axis and $y$-axis motion are presented separately, and frequently differ by $\sim$10-30$\%$ in the values for $R$ and $M$ determined for each axis.  From the data, it is easy to distinguish the high density Au particles from much lower density polystyrene nanospheres with comparable mass and brightness.  For the polystyrene, data from spheres of differing specified diameter (50-80 nm in 10 nm increments) are non-overlapping for all except one data point. 

The spread of our data for Au nanospheres substantially exceeds that quoted in the manufacturer's specifications.  In other experiments, we have made many accurate determinations of the mass of Au particles from the same manufacturer\cite{Cytodiagnostics}, using single charge stepping of $\CMR$ to determine $M$. From this data, we believe it is likely that the actual spread of the sizes of the Au particles exceeds the manufacturer's specifications.

Although the best data can only be obtained a day or so after collection, rapid (<1 hour) measurements are still possible and are valuable for quick characterization of collected particles: first, to distinguish high density (like Au) and much lower density organic contaminant particles. Secondly, aggregates of >1 particle will be distinguishable from single particles, since even quick data collection is unlikely to lead to mass errors exceeding 100$\%$.

\section{Transfer}
\label{sec:transfer}

There are several reasons to transfer a particle  from a collection trap to a different analysis trap for additional measurements, an operation we have previously described  \cite{Coppock2017}. First, as we emphasized in the previous section, charged particle collection can create stray electric fields that decay very slowly with time and can have a deleterious effect on measurements. Although this problem is likely to be most acute for particle collection using electrospray, it may also be relevant for other techniques\cite{Das2025}, such as laser-induced acoustic desorption \cite{Bykov2019}, where particles intended for collection may also accumulate on electrode surfaces. Particle transfer allows the particle to be moved out of the low vacuum collection chamber and into a chamber capable of much better vacuum\cite{Mestres2015}. The separate analysis trap and chamber can be configured and optimized for operations that are not practical in the collection chamber, such as exposure to an electron beam or deposition of trapped particles onto a substrate \cite{Coppock2024}. Indeed, once transfer technology is developed, it is possible to transfer particles between several traps, each optimized for different purposes.

\begin{figure}
\includegraphics[scale=1.00,draft=false]{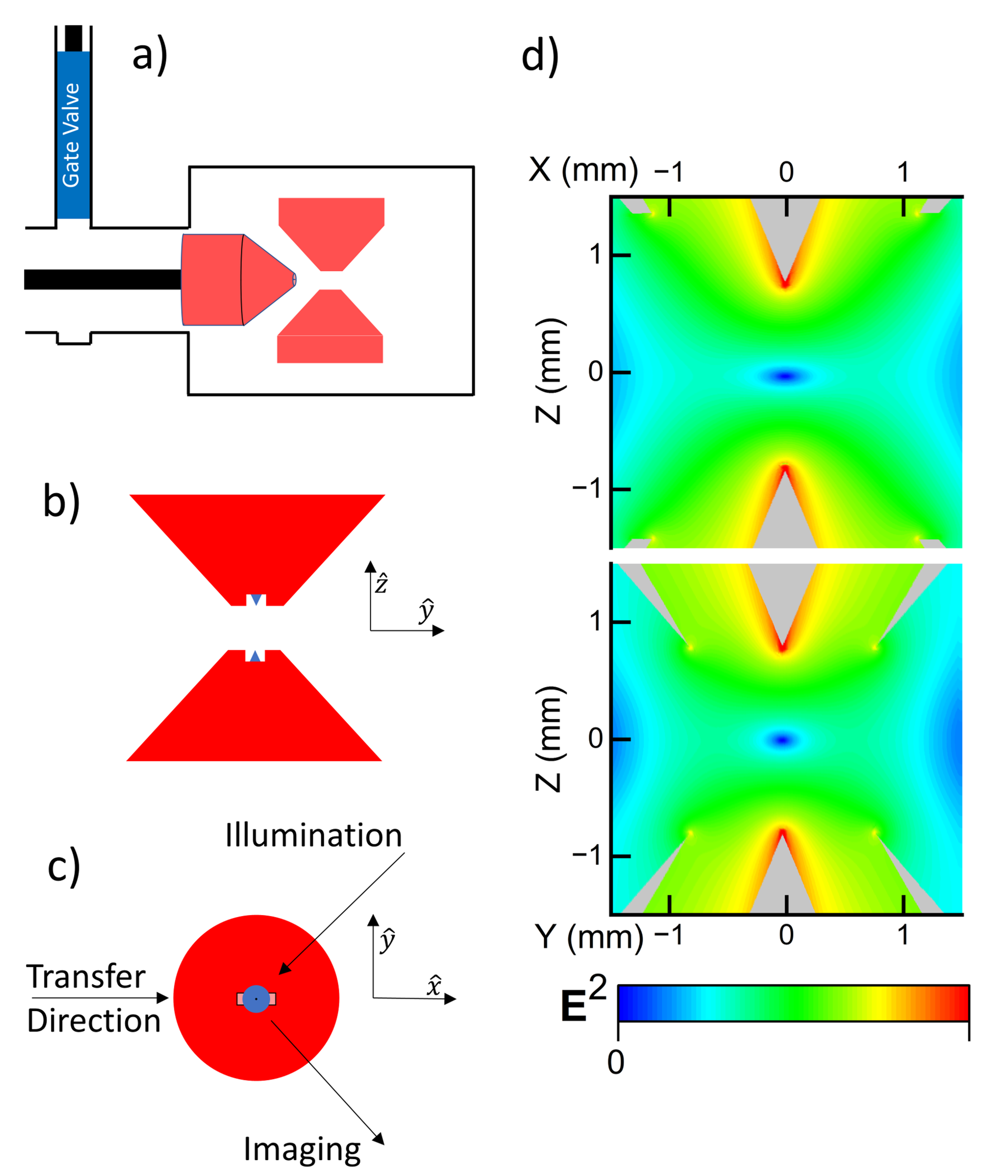} 
\caption{
(a) The trap in which the particle is collected is attached to a linear shift mechanism that allows it to be moved into close proximity to the biconal trap in the analysis chamber. The analysis trap is slotted (b) so that oscillatory motion in the trap occurs at three distinct frequencies that can all be observed by imaging optics located on an axis at angle 45$\degree$ away from the slot orientation (c). (d) COMSOL modeling of the slotted trap design, illustrating the different confinement potentials along and across the slot direction. $\Vac$ is applied to the outer electrodes, while both inner electrodes are grounded.
}
\label{Transfer}
\end{figure}

One highly attractive analysis trap design is the "biconal trap", illustrated in Fig. \ref{COMSOL}c and d, which is similar to designs previously used for single ion trapping \cite{NisbetJones2016,Lindvall2022} and for small particle analysis \cite{Esser2019}. It is capable of operation with the ac field applied either to the outer or inner electrodes, although outer electrode bias leads an additional "halo trap" surrounding the central trap (Fig. \ref{COMSOL}c).  Unlike the case for single cone traps, biconal trap potentials are symmetric, so odd terms in the polynomial expansion around the minimum are absent.  Biconal traps maintain ease of access for measurements at all angles in the  plane of symmetry once the collection trap has been retracted, an attractive feature when several probe beams, as well as particle deposition experiments, are contemplated.

A schematic depiction of the analysis trap and chamber is shown in Fig. \ref{Transfer}a. The collection and analysis chamber are separated by a gate valve. The collection (traveling) trap is attached to a 300 mm linear shift mechanism capable of alignment to $\sim$0.3 mm \cite{Shift} when fully extended. The outer electrode cone included angle for both the collection and analysis traps is 82$\degree$. The diameter of the end face of the conical outer electrodes is 1.59 mm, as is the separation between the opposing trap faces of the analysis trap. The inner electrodes, also conical  with an included  angle of 45$\degree$, have their apex in the plane of the trap faces. As is the case with our single cone  trap designs \cite{Nagornykh2015}, a slot is machined across the opposing analysis trap faces (width$=$1.20 mm, depth$=$0.60 mm). The slot intensionally breaks the rotational symmetry of the traps, so  that motion in the directions parallel and perpendicular to the slots have different frequencies. When the trap is viewed from a 45$\degree$ angle from the slot direction, motion along all three trap axes can be distinguished and analyzed (Fig. \ref{Transfer}b-d).

\begin{figure}
\includegraphics[scale=1.0,draft=false]{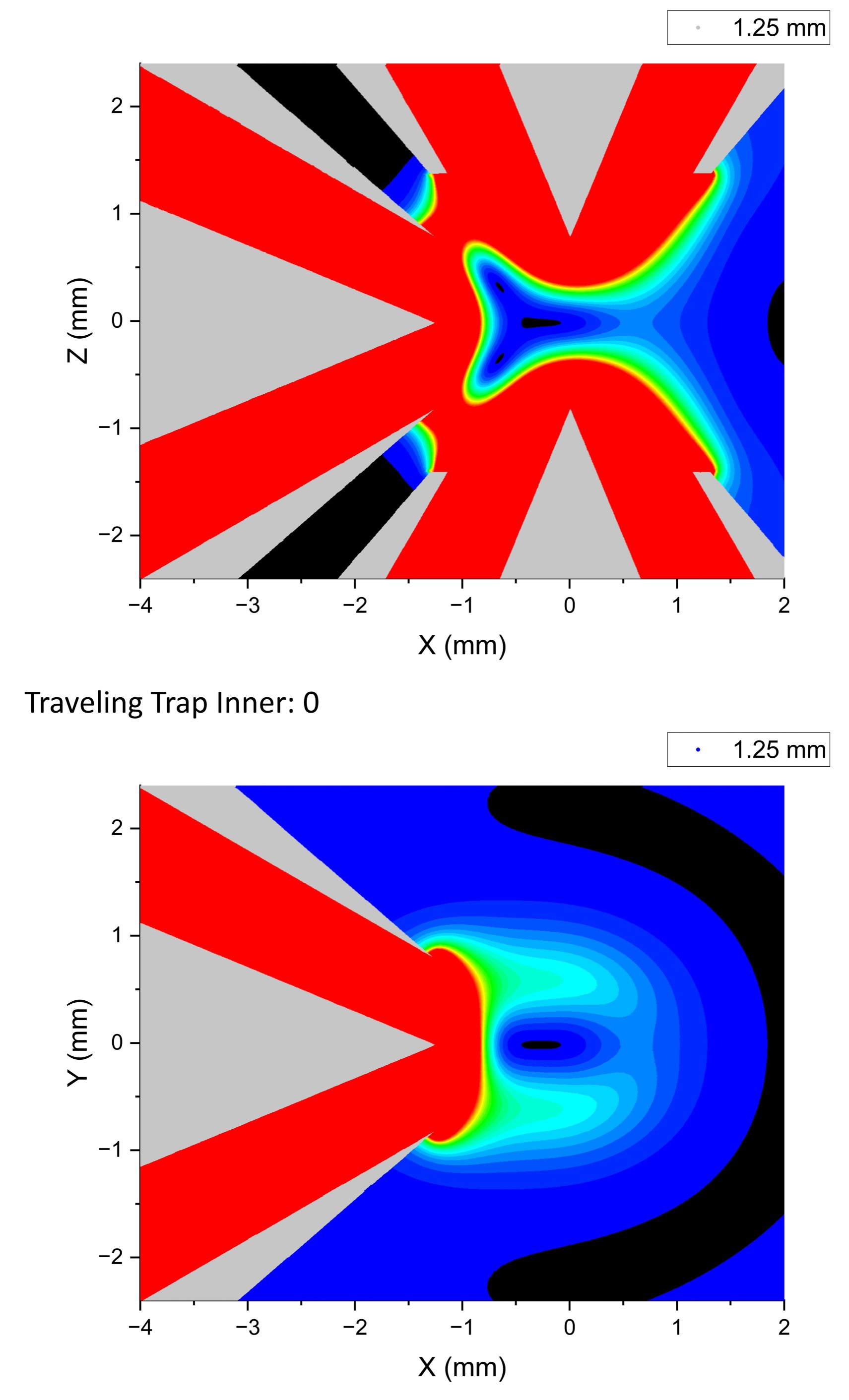} 
\caption{
(Multimedia available online) $\Psi$ (Eq. \ref{Psi}), plotted for the particle transfer process using COMSOL modeling, in the $x-z$ (top) and $x-y$ (bottom) plane.  An offset is introduced so that $\Psi=0$ at the trap minimum.  The color map is the same as in Fig. \ref{Transfer} but with black indicating $\Psi\cong 0$ and $\Psi<0$. Still image shows traps at closest approach (apex of traveling trap 1.25 mm from central axis of fixed trap) with $\Vdc=0$ on the traveling trap inner electrode.
}
\label{SlotMovie}
\end{figure}

For the transfer procedure, we apply the same ac voltage to the outer electrodes of both the collection and the analysis traps. After collection of the particle, a strong negative dc bias (typically -10 V) is applied to the inner electrode of the collection trap, drawing the particle close to the trap face.  Then, over a period of about 15 minutes, it is moved out of the collection zone and into transfer position, where the face of the collection trap is a distance $\cong$1.25 mm from the center of the analysis trap (Fig. \ref{SlotMovie}. Multimedia available online). The dc bias of the analysis trap inner electrodes is maintained at ground throughout the transfer process. Transfer is accomplished by ramping the bias on the inner electrode of the collection trap to a large positive potential (typically +10 V), pushing the particle into the analysis trap. The collection trap is then retracted with its inner electrode bias potential held at the large positive value.

\begin{figure}
\includegraphics[scale=1.0,draft=false]{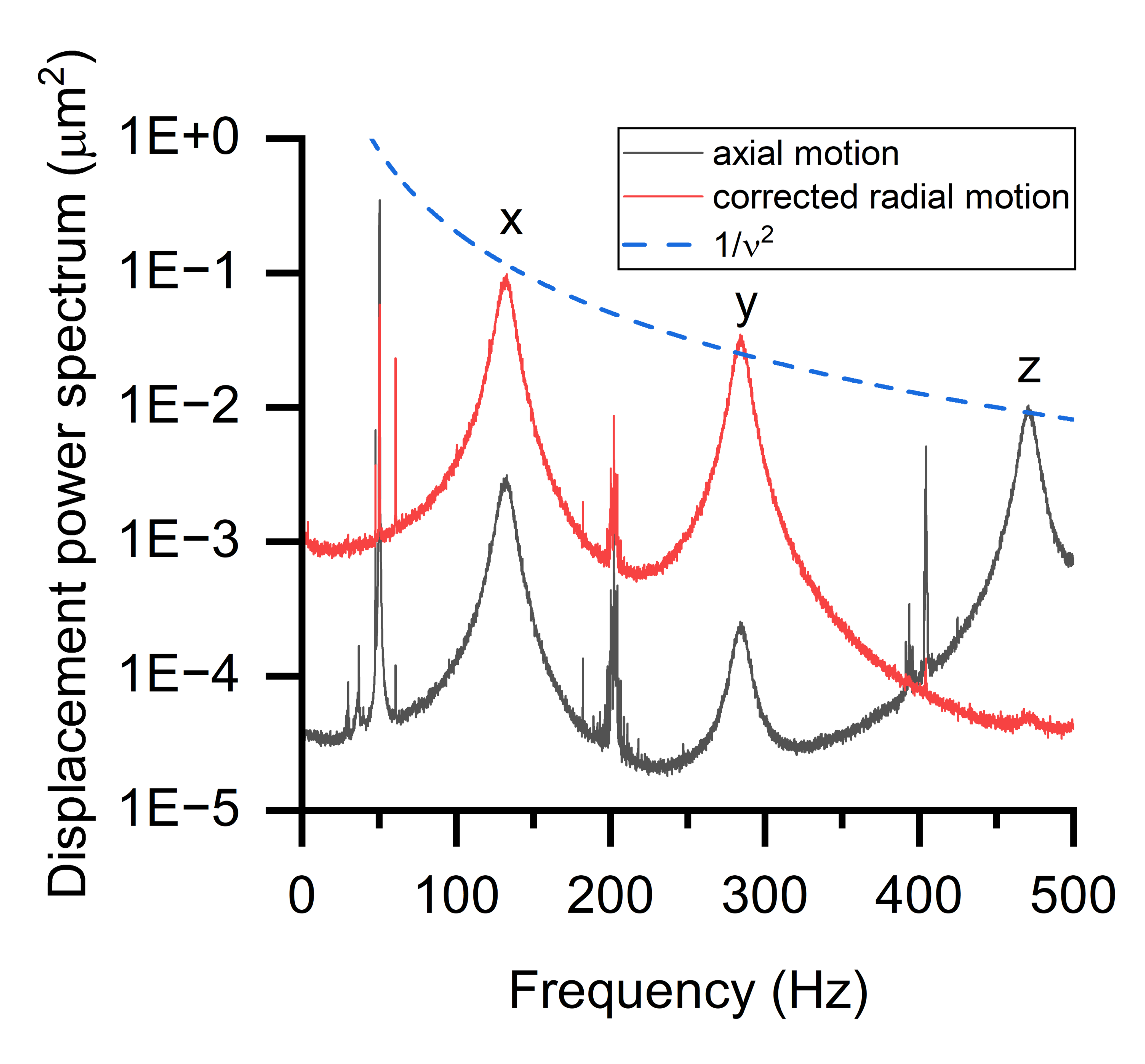} 
\caption{
Spectra of motion of a particle transferred to the analysis trap.  Three frequencies appear in the spectra because of the slotted trap geometry and the direction relative to the slot of the imaging optics (see Fig. \ref{Transfer}). Rescaling of the radial motion spectrum is necessary to account for the fact that the axes of motion are not perpendicular to the image plane. Dashed line indicates expected peak amplitudes for thermal motion (Eq. \ref{Eq. e}). Sharp peaks are caused by pumps and instrumentation. 
}
\label{ThreeAxis}
\end{figure}

To demonstrate transfer,  we collected a 600 nm polystyrene particle with $\CMR\simeq7~ \cmrunits$ in the collection trap. We transferred the particle at $p=$25 mTorr, using $\VAC=$300 V at $\Omega/2\pi\simeq$20 kHz on all trap outer electrodes. After transfer and retraction, the particle is  observable with an additional camera  capable of fast measurements. The spectrum of motion using this imaging camera is shown in Fig. \ref{ThreeAxis}, and it exhibits the distinctive characteristics of motion in the asymmetric analysis trap.  By reversing the order of the process steps, we are also able to return the particle from the analysis trap back to  its original position in the collection chamber.

\section{Particle position sensing and control techniques for precision measurements and high stability}
\label{sec:Control}

\begin{figure*}
    \centering
    \includegraphics[width=1\linewidth]{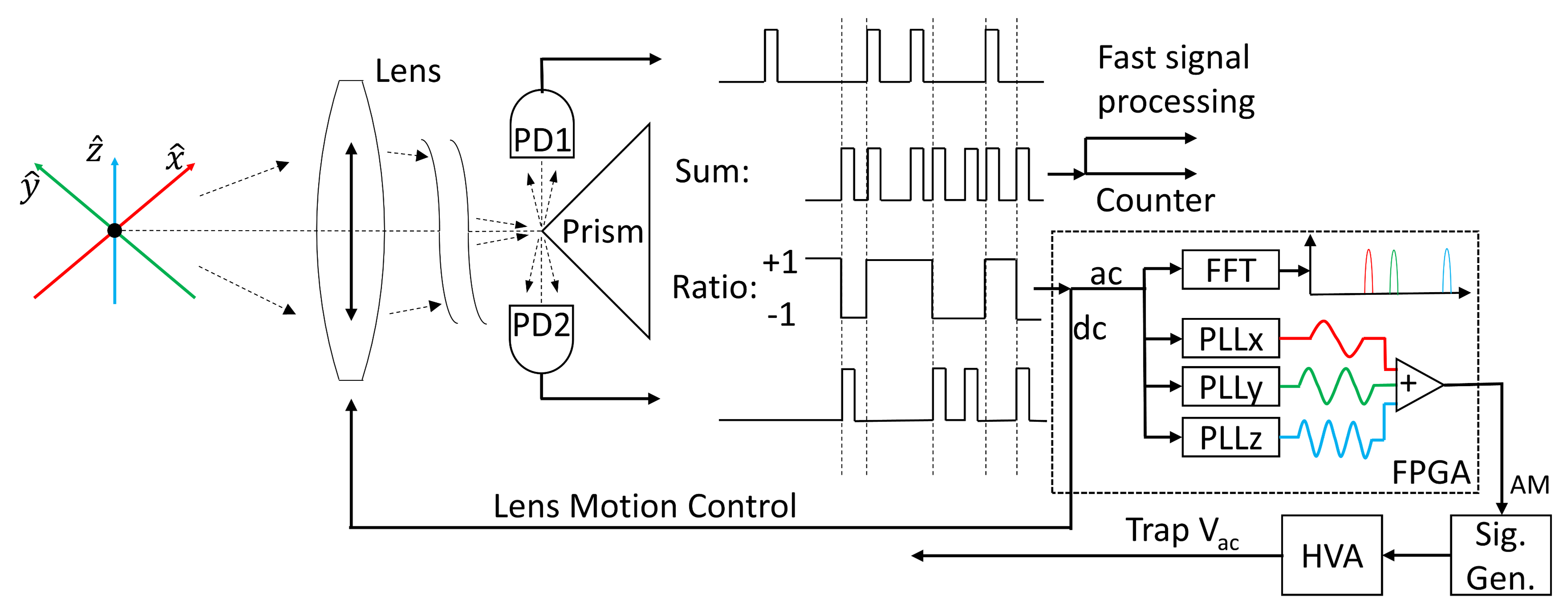}
    \caption{Schematic  diagram of the optics and electronics used to measure and control oscillatory motion of a trapped particle.  A lens (located outside the vacuum chamber) focuses the image of the particle onto a knife edge prism reflector.  The optics are oriented so that motion in the $x$,$y$, and $z$ directions each have equal projections on the image plane in the direction perpendicular to the knife edge (vertical in the figure). Light reflected off of the knife edge is collected by two avalanche photodetectors, and their output pulses are either summed or ratioed by pulse electronics. The ratio signal is sent to a controller with a field programmable gate array (FPGA)\cite{NI} that performs three tasks: (1) The dc signal is used to control a stage where the lens is mounted, so that the particle image remains focused on the knife edge. (2) An FFT (0.1 Hz-3125 Hz) is performed to track the frequencies of the particle's oscillation. (3) Three phase locked loops (PLLs) track the particle's oscillations and are used to control their amplitude via parametric feedback cooling and heating.
}
    \label{Pulses}
\end{figure*}

Once a particle has been transferred into an analysis chamber and pumped to high vacuum, a wealth of information about the particles's properties can be  gleaned from precision measurements of $\CMR$. Since $Q$ can be determined from charge quantization, $\CMR$ measurements enable mass determination: mass loss can be an indication of evaporation or sublimation of the trapped particle\cite{Howder2015,Rodriguez2022,Friese2025}, while mass gain can be a sensitive probe of surface adsorption from residual gases in the chamber\cite{Ricci2022}\cite{Hoffman2020}. High vacuum measurements necessitate an instrumental mechanism to cool the particle, since friction from background gases becomes negligible, and heating mechanisms (like those that impacted the data in Section \ref{sec:characterization}) are inevitably present.

We have previously described our techniques for obtaining long term particle stability for measurements at $p$ below $10^{-6}$ Torr\cite{Nagornykh2015}.  We employ  a knife edge beam splitter and two small arrays of avalanche photodetectors\cite{MPPC} to sense motion along all three axes of a particle confined in an ion trap (Fig. \ref{Pulses}). Compared to the camera technique described in Section \ref{CollectionApparatus}, the photodetectors have a faster response time, which is relevant both for position detection and for rapid measurements of particle rotation\cite{Nagornykh2017}. The output pulses from the detectors are summed and counted to determine the total scattered light from the particle. Additionally, they are fed to a circuit that switches between two possible values: -1 V and +1 V. When its value is at -1 V, it switches to +1 V only when a pulse on PD1 arrives; when its value is at +1 V, it switches to -1 V only when a pulse on PD2 arrives (Fig. \ref{Pulses}). Such a circuit can be shown to output an average value of $(I_1-I_2)/(I_1+I_2)$, where $I_1$ and $I_2$ are the average pulse rates of the two photodetectors. The filtered dc value of this ratio signal is used to keep the image of the  particle always focused on the knife edge, while its ac component is used to analyze and control the particle's oscillations in the trap potential.

\begin{figure}
\includegraphics[scale=1.0,draft=false]{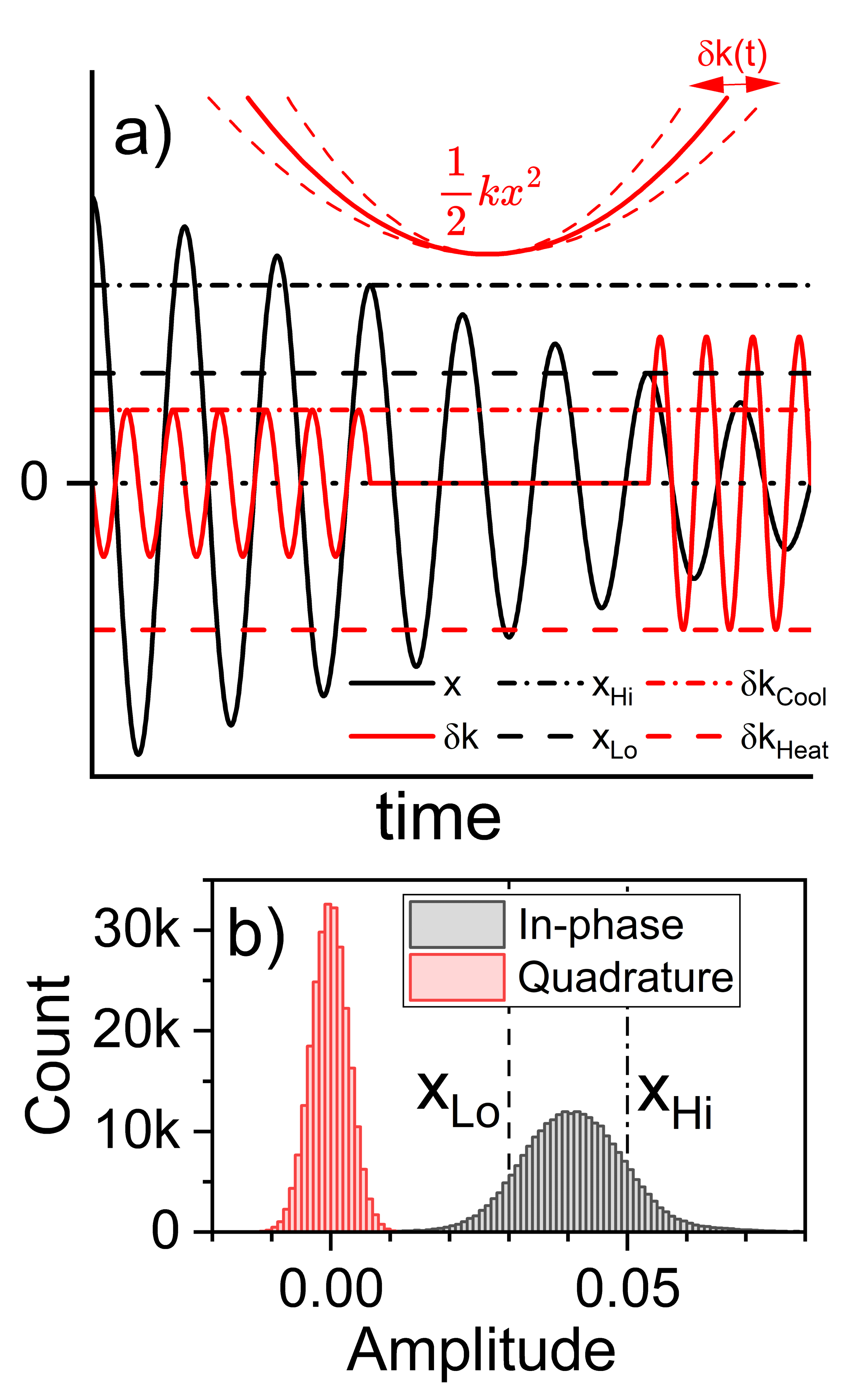} 
\caption{
(a) Schematic depiction of the principle of parametric heating and cooling: for a particle oscillating in a potential $U(x,t)=k(t)x^2/2$ with motion in phase with $\cos \omega t$, a modulation of $k$ with $\delta k(t) \propto -\sin 2 \omega t$ will reduce the kinetic energy of the particle, while a modulation of the opposite sign will increase it. In our experiments, this modulation is generated by a phase locked loop that tracks particle oscillations, and is used to modulate the amplitude $\Vac$ of the ion trap potential. When the amplitude of particle oscillation exceeds $x_{Hi}$, modulation is set to cool the  particle, while when the amplitude is less than $x_{Lo}$, the particle  is heated.  No modulation is applied when particle motion amplitude is within the set bands. (b) Histogram of the in-phase and quadrature magnitudes of a trapped particle's motion, determined by multiplying the measured signal by the PLL output and integrating the result over a period $\cong$1.3 s. Data was collected for $\simeq$24 hours on the same particle that was used for the data in Fig. \ref{NoisePlot}a.
}
\label{Parametric}
\end{figure}

In trap designs with a slot geometry, as in Fig. \ref{Transfer}, motion within the trap has three distinct frequencies.  To control motion in all three frequencies we employ three phase locked loops (PLLs) that track motion in all directions of the ion trap. We use parametric feedback cooling (Fig. \ref{Parametric}) to stabilize the particle in high vacuum\cite{Nagornykh2015}.  Appropriately phased signals at twice the particle oscillation frequencies are summed and used to modulate the amplitude of the trap voltage $\VAC$.

The simplest possible cooling algorithm, where the parametric feedback phase is set for maximum cooling, has proven adequate for long term stability ($\gtrsim$weeks), with center of mass  temperatures of 1-10 K at $p\leq 10^{-6}$ Torr. However, measurements of trap oscillation frequencies (from which one can deduce $\CMR$) are best performed when the oscillation amplitudes are not too small. We have also found that the simple cooling algorithm  often fails at $p\leq 10^{-7}$ Torr, when the amplitude of oscillations can become so small that the PLL is unable to  track the signal.  When this happens, the particle is not lost, but $\CMR$ measurements are compromised.

To remedy this problem, we have added thermostatic control to our cooling algorithm (Fig. \ref{Parametric}). The amplitude of signal oscillations is tracked by measuring the integral of the product of the signal, $x(t)$, and PLL output tracking the signal, $\mathrm{Cos~\omega_x t}$. An output proportional to $\mathrm{Sin~2\omega_x t}$ (with a  negative sign for cooling and a positive sign for heating) is summed with analogous outputs for the motion of $y$ and $z$ to modulate the amplitude of the trap $\VAC$.  Our implementation maintains oscillation amplitudes within a band: above an upper limit, feedback provides cooling, while below a  lower limit, feedback heats the particle.  All feedback is suppressed within the  range of the lower and upper bands.

\begin{figure}
\includegraphics[scale=1.0,draft=false]{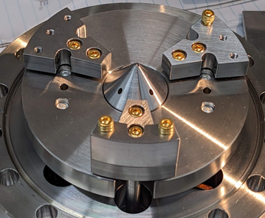} 
\caption{
Photograph of the trap used for precision measurements of $\CMR$ in high vacuum.  Trap geometry is identical to that discussed in Section \ref{sec:transfer} and Fig. \ref{Transfer}, except that the trap consists of a  single  cone with an outer and inner electrode.  The Trap is surrounded by a cylindrical grounded plate and by three electrodes spaced 120$\degree$ apart that are biased to counteract dc electric fields at the trap center. All electrodes are made from stainless steel.
}
\label{TrapPic}
\end{figure}

A photograph of the trap used in the experiments described in the following sections is shown in Fig. \ref{TrapPic}. We have used this single cone design in our previous experiments, and it is incorporated into a system with particle transfer capabilities that can be operated at $p\sim10^{-8}$ torr.  Stray dc electric fields at the trap center are nulled using three electrodes surrounding the trap. Digital noise in a narrow band around each of the three frequencies of motion in the trap is added to the modulation input of the signal generator, and the correlated motion of the particle is minimized by feedback to the electrodes\cite{Eltony2013,Nadlinger2021}.

\section{Precision measurements of $\CMR$ in high vacuum}
\label{sec:precision}

To test the effectiveness of thermostatic feedback  and to assess the potential for high precision measurements enabled by this technique in high vacuum, we use Au nanospheres similar to those we have worked with previously\cite{Coppock2021,Coppock2022}, with masses around $1\times10^{-16}~$kg and typical $\CMR\cong$1-3 $\cmrunits$. We set the values for the band limits to maintain an amplitude of $\sim2~\mu$m for motion along each of the three axes, which is comparable to the thermal motion expected at $T\sim$100 K.  We have operated the thermostatic control algorithm successfully in the range $10^{-8}\leq p\leq10^{-5}~$Torr.  At the lowest pressures, measurements can remain stable--all the PLLs are locked to the particle motion--for several weeks. 

While PLLs are used stabilize the oscillation amplitudes, precision measurements are made with separate FFT analysis.  The ratio signal is sampled at 6250 Hz, and $2^{16}$ data points are used so that a spectrum is produced with $\simeq$0.1 Hz spacing from dc to 3125 Hz.  Data in the FFT (Hanning) window is updated and a new spectrum is obtained every $\cong$1.3 s
(see Ref.\citenum{FPGA}). 

When $p\leq10^{-5}~$Torr, the linewidth of the peaks is $\ll$0.1 Hz, but the peak amplitudes differ  from the noise  in more than one element of the spectrum, due to the finite duration of the sampling window.  To estimate the location of the oscillation frequency, we perform an average of the frequencies, weighted by their amplitudes, over a region of 10 elements surrounding the peak of maximum amplitude.  In this way, we can determine the location of the oscillation frequency to a higher resolution than the 0.1 Hz spacing between frequencies of the FFT.

Particle motion in an ion trap for small oscillations about the trap center ($x,y,z=$0) is governed by the Mathieu equation. For motion in a  single dimension:

\begin{equation} \label{Eq. g}
\frac{d^2 z}{d\tau^2} + \left[a_z - 2q_z\cos(2\tau)\right]z = 0,
\end{equation}
where $\tau=\Omega t/2$,

\begin{equation}
\label{Eq. h}
a_z=4\,\frac{Q}{M\Omega^2}\partial_z^2 U_{dc}\big|_{\mathbf{z}=0}
\quad\text{and}\quad
q_z=2\,\frac{Q}{M\Omega^2}\partial_z^2 U_{ac}\big|_{\mathbf{z}=0}.
\end{equation}
$U_{ac}$ is the potential induced  by the voltage, $V_{ac}$, applied to the trap. $\partial_z^2 U_{ac}\big|_{\mathbf{z}=0}=V_{ac}/z_0^2$, where $z_0$ is a distance characteristic of the trap geometry. $U_{dc}$ arises from accumulated charge or patch potentials on the trap surface.  Note that the feedback used to eliminate dc electric fields at the center of the trap cannot eliminate second derivatives of $U_{dc}$.

For some values of $a_z$ and $q_z$, Eq. \ref{Eq. g} has stable solutions with well-defined characteristic oscillation frequencies, $\omega_z$.  For $q_z \ll1$ and $a_z>0$, these frequencies are given by: 

\begin{equation} \label{Eq. i}
\beta_z \equiv\frac{2 \omega_z}{\Omega} \cong \sqrt{a_z+\frac{q_z^2}{2}}\quad \mathrm{and:}\quad
|q_z|(a_z=0)=\sqrt{2}\,\beta_z.
\end{equation}
When $a_z \rightarrow0$, the second equation can be used to determine $\CMR$ from measurement of $\omega_z$:

\begin{equation} \label{Eq. j}
\CMR=\frac{q_z \Omega^2 z_0^2}{2 V_{ac}}\cong\frac{\sqrt{2} \omega_z \Omega z_0^2}{V_{ac}}.
\end{equation}
Similar equations apply for motion along the $x$ and $y$ axes, and we will drop the subscripts below when this fact is applicable.

In general, dc electric fields and gradients are non negligible, and--since the values of $a_x$,$a_y$, and $a_z$ will likely arise from random patch potentials on the trap electrodes--there is no way we can determine or eliminate these terms from frequency measurements alone \cite{Lindvall2022}. Instead, we proceed by assessing the importance of dc gradients to our measurements by using oscillation frequency data to determine: $\gamma\equiv(q_z-q_y-q_x)/(q_z+q_y+q_x)$. As long as $\omega_z>\omega_y,\omega_x$, $q_z=q_x+q_y$ by Laplace's equation and $\gamma$ is zero.  Deviations from zero are an indication that dc fields are having an impact on the values of the oscillation frequencies.

Because Eq. \ref{Eq. i} is approximate, for more precise measurements, we use:

\begin{equation} \label{Eq. k}
\begin{aligned}
|q|(a=0)
={}&\sqrt{2}\,\beta\bigl(
1-0.390627\,\beta^2+0.026696\,\beta^4 \\
&\hphantom{\sqrt{2}\,\beta\bigl(}
+0.007901\,\beta^6-0.001886\,\beta^8
\bigr).
\end{aligned}
\end{equation}
This series, whose terms were found numerically using  Mathematica\cite{Mathieu} has error of less than $10^{-6}$ for the full range  of stable solutions with $a=0$ and $0<q<0.9$

As can be seen from Eq. \ref{Eq. j}, both $\Omega$ and $\Vac$ contribute to $\CMR$.   In our experiments, $\Omega$ is sourced from a signal generator with a quoted frequency stability of 1 ppm. $\Vac$, however, is the output of a high voltage amplifier that has significant ($\sim$0.1$\%$) long term amplitude variations, probably due to temperature fluctuations. To account for these fluctuations, we measure the amplitude with a lock-in amplifier with the OFFSET/EXPAND feature activated, thus eliminating digitization errors.  This lock-in output is then used to determine the value of $\Vac$ to use in Eq. \ref{Eq. j} to calculate $\CMR$. The geometric trap parameter $z_0 \cong$1.99 mm is determined by ejecting particles from the trap and measuring time of flight to a charge sensing substrate \cite{Coppock2024}.

 Because Au is highly absorptive, illumination power must be well below that where significant evaporation can occur: we use a 532 nm laser with a power density of $\simeq$1000 $\Wpermm$.  This level of illumination produces $5-10\times10^5$ counts$\cdot \persec$ on the sum output of the signal processing circuit (Fig. \ref{Pulses}). Measurements are performed in a turbo pumped  vacuum chamber with $p\simeq2\times10^{-8}~$Torr.

\begin{figure}
\includegraphics[scale=1.0,draft=false]{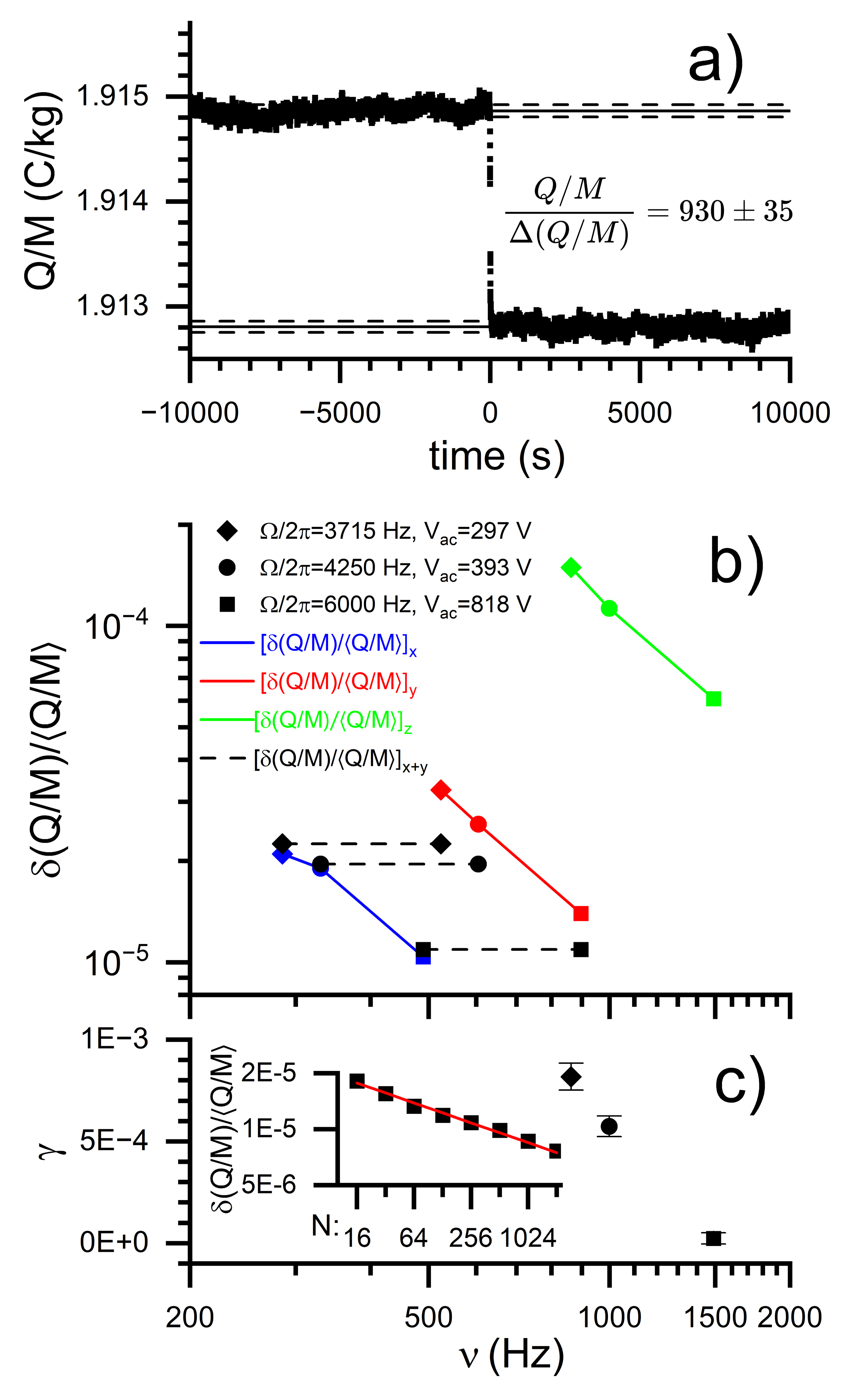} 
\caption{
(a) Precision measurement of $\CMR$ of an Au nanosphere, taken with $\nu_{ac}$=5000 Hz and $\Vac$=408 V, showing data of the loss of a single elementary charge. Data is taken at 1.3 s intervals.  (b) Values of $\delta(Q/M)/\langle Q/M \rangle$ derived using measurements of only a single frequency, either $\nu_x$,$\nu_y$, or $\nu_z$, plotted for three different trap bias parameters for each frequency. Data taken at 1.3 s intervals is averaged in 256 data point blocks. Statistics are derived from a $\sim$24 hour data set of these block averages during a period when the particle $Q$ remained constant. Because data derived from $\nu_z$ is more noisy, our usual determination of $\CMR$ (horizontal dashed lines) relies exclusively on  measurements of $\nu_x$ and $\nu_y$. (c) Data for $\gamma\equiv(q_z-q_y-q_x)/(q_z+q_y+q_x)$, plotted as a function of $\nu_z$. (Inset) Data (for $\Omega/2\pi$=6000 Hz and $\Vac$=818 V) for $\delta(Q/M)/\langle Q/M \rangle$, plotted as a  function of block size, $N$. The red line is a power law fit with $N^{-1/6}$.
}
\label{NoisePlot}
\end{figure}

$\CMR$ data is shown in Fig. \ref{NoisePlot}a near a time when $Q$ changes by a single elementary charge.  Data points are plotted at intervals of 1.3 s, corresponding to the updates of FFT data. While the charge  loss event is likely instantaneous at these time scales, the data makes a complete transition across the step in about 30 s. The  single  discharge  event allows the mass  to be  determined: $M=7.1\pm0.25\times10^{-17}$ kg.

Since we track $\nu_x$, $\nu_y$, and $\nu_z$ separately, it is possible to use any subset of them to estimate $\CMR$.  In Fig. \ref{NoisePlot}b, we plot the ratio of the standard deviation to the mean, $\delta(Q/M)/\langle Q/M \rangle$, determined from each of the three modes of oscillation.  Each $\CMR$ data point is an average of 256 data points taken at 1.3 s intervals and is taken continuously over a $\sim$24 hour period where there are no discharge events. Data is also collected for three different sets of trap parameters.

Two facts about the data are important: first, data derived from $\nu_z$ has significantly more noise than the other two modes of oscillation. Consequently, we exclusively use  $q_x+q_y$ to determine $\CMR$ (as in Fig. \ref{NoisePlot}a). Second, for all three modes of oscillation, the noise diminishes as the trap parameters are adjusted to increase their frequencies.
These observations are consistent with the hypothesis that low frequency fluctuations in the dc electric fields associated with charge on the trap electrodes are the primary source of noise, and that fields along the trap $z$ axis have the largest magnitudes.  This idea is corroborated by the behavior of $\gamma$ shown in Fig. \ref{NoisePlot}c, which rapidly declines to $<10^{-4}$ when the trap frequencies used to determine it are at their highest values.

$\delta(Q/M)/\langle Q/M \rangle$ has a very modest dependence on the number of adjacent data points, $N$, used in the average (see Fig. \ref{NoisePlot}c (inset)).  This means that it is possible to take high precision data ranging from the $\sim$30 s response time indicated in Fig. \ref{NoisePlot}a to long data sets encompassing hours of averaging taken over many days. For averaging times greater than $\sim$ 10 minutes, $\delta(Q/M)/\langle Q/M \rangle < 10^{-5}$.

Finally, to determine $precisely$ the number of charges, $n$, on a particle by measuring the $\CMR$ difference across $m$ consecutive discharge  (or charge) steps, it can be shown that:

\begin{equation} \label{Eq. l}
\frac{n^2\delta(Q/M)}{m\langle Q/M \rangle}<1,
\end{equation}
assuming that $M$ is unchanging during the measurements. Thus, the precision of $\delta(Q/M)/\langle Q/M \rangle\simeq10^{-5}$ is insufficient to determine exactly the charge on the object in Fig. \ref{NoisePlot}a, but a single step determination of the exact charge would be possible for $n\leq300$. 

\section{Conclusion}
\label{sec:conclusion}

We have presented techniques and methods for characterization and precision measurements of nanoscale particles confined in an ion trap.  Imaging of thermal motion of trapped particles is a fast way to characterize the mass and size of particles, which is useful in situations where there is uncertainty in what is  collected (possible particle aggregation, for example).  This data has also provided insight into non-thermal noise that is relevant when ion traps are exposed to charged particles.  Our transfer technique may have many applications in situations where the optimal measurement environment may not be suitable for particle collection. This technique could in principle be extended to transferring a particle between several traps, each optimized for  specific tasks or measurements.

Interestingly, we have found that once a particle has been transferred to an analysis trap and pumped to high vacuum,  all feedback control can be interrupted and illumination of the particle entirely turned off for periods of up to 24 hours without particle loss from the trap. The ability to do this may have important applications for investigations of levitated materials where light or thermal effects have an effect on material properties.

Although our techniques for precision measurements have been developed and proven at $p\ge10^{-8}$ Torr, we expect that they will also be applicable in the ultra high vacuum (UHV) regime. However, it will be necessary to carefully shield the trap neighborhood from charges emanating from ion pumps or other sources of ions and electrons to prevent unwanted electric fields and gradients from affecting trap measurements.

Particle stabilization and precision measurements in high vacuum will open the door to new probes of nanoscale materials.  For example, the precision we report of $\delta(Q/M)/\langle Q/M \rangle\simeq10^{-5}$ should allow measurements of submonolayer surface accumulations on micron-scale particles in UHV, a potentially valuable tool for surface chemistry.  These techniques, in combination with  precision rotation \cite{Nagornykh2017} and controlled deposition of levitated particles \cite{Coppock2024}, may also enable the growth and characterization of new materials developed in the levitated environment.

%

\begin{acknowledgments}
This work was supported by the Laboratory for Physical Sciences, Contract \#H9823023C0086.
\end{acknowledgments}

\section*{Author declarations}

\subsection*{Conflict of Interest}

The authors have no conflicts to disclose.

\subsection*{Author Contributions}

\textbf{B. E. Kane}: Conceptualization, Methodology, Software, Formal analysis, Investigation, Visualization, Writing – original draft, Writing – review and editing.  \textbf{Joyce Coppock}: Software, Formal analysis, Supervision of data analysis, Writing – review and editing. \textbf{Sunghyun Kim}: Investigation, Writing – review and editing. \textbf{Sarah Westgate}: Investigation, Formal analysis, Visualization, Writing – review and editing. All authors reviewed and approved the final manuscript.

\section*{Data Availability}
The data that support the findings of
this study are available from the
corresponding author upon
request.

\bibliography{Techniques}

\end{document}